\documentclass[12pt]{iopart}
\usepackage{graphicx}

\begin{document}

\title[Brain, Rain and Forest Fires]{What is critical about criticality: In praise of the correlation function}

\author{Henrik Jeldtoft Jensen}

\address{Centre for Complexity Science and Department of Mathematics, Imperial College London, South Kensington Campus, SW7 2AZ, UK\\
and\\
Institute of Innovative Research, Tokyo Institute of Technology, 4259, Nagatsuta-cho, Yokohama 226-8502, Japan	\\
}
\ead{h.jensen@imperial.ac.uk}
\vspace{10pt}
\begin{indented}
\item[] 17 June 2021
\end{indented}

\begin{abstract}
We present a brief review of power laws and correlation functions as measures of criticality and the relation between them. By comparing phenomenology from rain, brain and the forest fire model we discuss the relevant features of self-organisation to the vicinity about a critical state. We conclude that organisation to a region of extended correlations and approximate power laws may be behaviour of interest shared between the three considered systems.  
\end{abstract}

%
\noindent{\it Keywords}: Rain, Brain, Forest Fire, Finite size criticality %
%
%
%

\section{Introduction}
The critical state and its vicinity have repeatedly attracted great interest probably starting from the study of the physics of second order phase transitions through out the 20th century which culminated in the renormalisation group theory\cite{Goldenfeld1992}. The notions such as ``Edge of Chaos''\cite{Packard1988} and and ``Self-Organised Criticality'' (SOC)\cite{BTW1987} from the 1980ties coming from different starting points, dynamical systems theory and high-dimensional non-equilibrium statistical mechanics, respectively, were inspired by the peculiar nature of the hyper sensitive critical state and sought mechanisms that would explain the seemingly omnipresence of this remarkable state. 

Although the reference to edge of chaos and criticality sometimes is use synonymously in a somewhat loose descriptive way\cite{Barras2013,StLouis2019}, both concepts have precise mathematical definitions and only occur if the control parameters of the system are tuned to certain values. 

A dynamical system is chaotic if it has a positive Lyapunov export\cite{Ott1993}. Chaotic behaviour can occur even when only a few parameters, usually denoted as degrees of freedom, are involved. Incase  the time evolution can be captured by an iterative map, as for example the population size of a species with non-over lapping generations, e.g. annual insects,  chaos can occur even for a single degree of freedom, $x_n$. The most famous is probably the logistic map\cite{May1976} given by the equation
\begin{equation}
	x_{n+1}=rx_n(1-x_n),
	\label{logistic}
\end{equation}
which evolves chaotically for the coefficient $r$ larger than $r_\infty\simeq 3.57$.

In contrast the critical state of equilibrium statistical mechanics can strictly speaking only occur in the limit of infinitely many degrees of freedom. The hallmark of the critical state is the long range correlations\cite{Goldenfeld1992}. The prototypic system supporting a critical state is the Ising model. In its basic version, the model consist of \textit{spin} variables $S_i\pm 1, i=1,2,...,N$ with $i$ labelling the sites on a $d$ dimensional hypercube of linear extension $L$, so $N=L^d$. The variance  coefficient
\begin{equation}
	C(r_{i,j})=\langle S_iS_j\rangle -\langle S_i\rangle\langle S_j\rangle,
	\label{Corr_func_1}
\end{equation} 
where $r_{ij}$ is the distance between site $i$ and site $j$, is taken as a measure of the interdependence between different locations. Although this object is denoted the variance in statistics, in the statistical mechanics literature it has always been denoted the correlations function. We will therefore stick to this tradition. The dependence on the separation $r=r_{ij}$ is found to be described by the following mathematical form
\begin{equation}
	C(r) \propto r^{-a} e^{-r/\xi}.
	\label{Corr_func_2}	
\end{equation}
 Here $\xi$ is called the correlation length and is a finite length scale except in the critical state. The critical state is reached by tuning the temperature $T$ to a specific value denoted the critical temperature $T_c$. As $T$ approaches $T_c$ the correlation length goes to infinity $\xi(T)\rightarrow\infty$. 
 
 We notice that for $T\neq T_c$, the correlation function will decay algebraically, i.e. as a power law $C(r)\sim r^{-a}$ for length scales short compared to $\xi$, i.e. $r \ll \xi$ and exponentially, $C(r)\sim e^{-r/\xi}$  for large separations $r\gg \xi$. 
 
Only exactly in the critical state $T=T_c$ will the correlations decay slowly as a power law $C(r)\sim r^{-a}$ no matter how large the separation $r$ and and since $\xi\rightarrow\infty$ no single characteristic scale distinguish the spatial and temporal organisation. We say the state is scale invariant because it looks and behaves in the same manner at all length and time scales except for a magnification factor.  As a consequence spatial structures are expected to take the form of  fractals. And probability distribution, such as those describing the response to external perturbations, are expected to follow power laws.

The suggestion of SOC\cite{BTW1987,Jensen1998,Pruessner2012} is that in very many systems out of equilibrium, the dynamics, when slowly driven, will self-organise to a state critical in the sense just described. The philosophy of the original paper\cite{BTW1987} was that since the critical state is scale invariant, criticality can be studied by investigating the probability distributions describing the abrupt releases of pent-up strain as a consequence of gradually added stress. If such distributions follow power laws, we expect to be dealing with a critical state, even if we are unable to check directly that the correlation function in Eq. (\ref{Corr_func_1}) does indeed depend on separation algebraically. Although this is a sensible starting point, it may not always be correct.

Distributions of event sizes, we will follow the tradition and use the generic term avalanches, can follow power laws even if the individual relaxation events generating the avalanche are entirely independent. The uncorrelated branching process and the independent random walk are representative examples.
\begin{figure}[!htbp]
        \centering\noindent
         \hspace{1.9cm}
       \includegraphics[width=13cm]{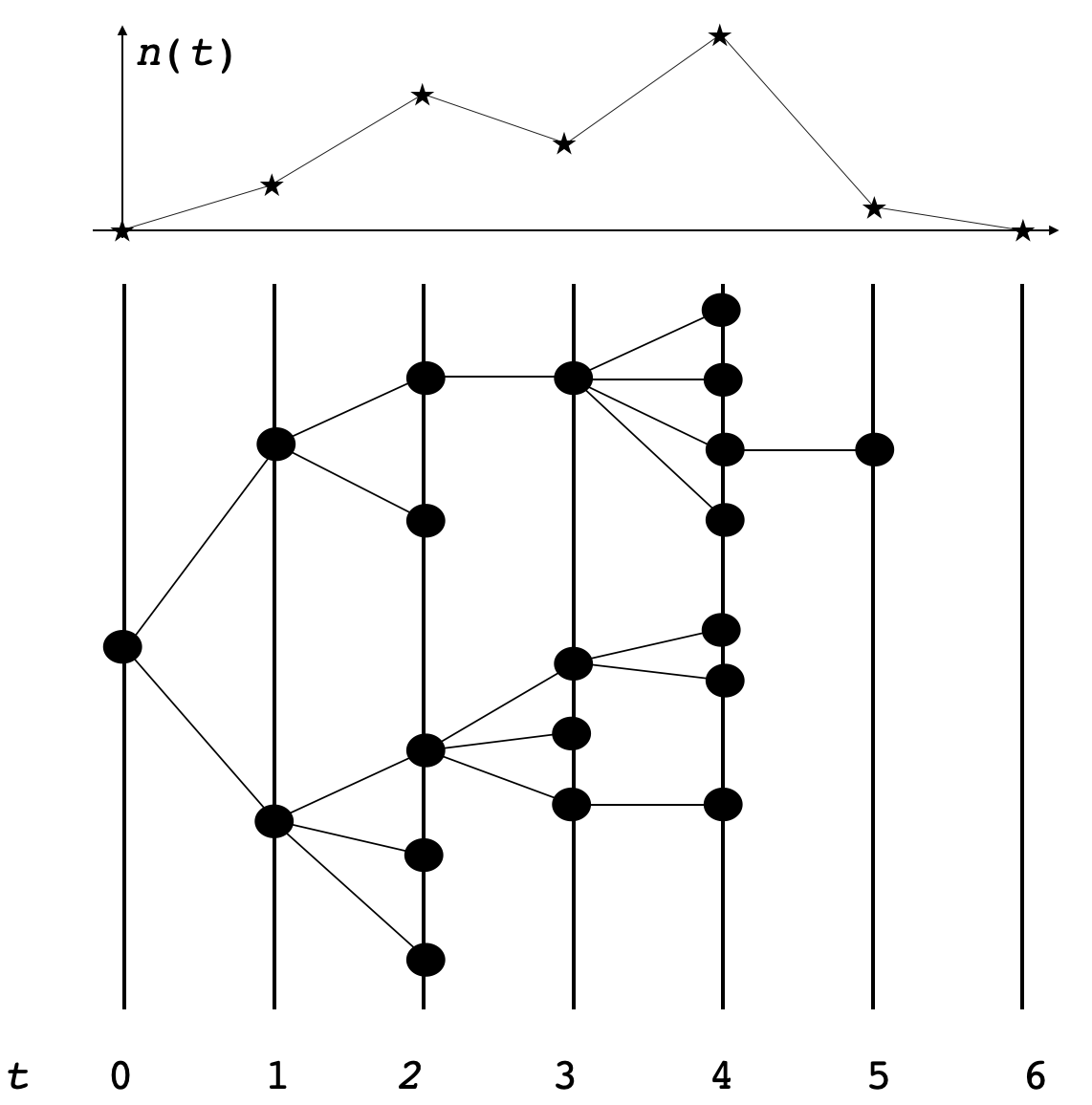}
       \caption{A realisation of a branching process. Each node produce $k$ offspring $k=0,1,2,...$ with the same independent probabilities $p_k$. Time is counted in generations $t$ and the evolution of the tree, or avalanche, is described by $n(t)$, the number of nodes  (see top panel) in generation $t$. The duration $T$ of the avalanche is from generation $t=0$ up to the last generation for which $n(t)>0$, so in the figure $T=6$.}
       \label{Branch_fig}
\end{figure}

Let us briefly recall the situation. An uncorrelated branching process consists of ``nodes'' producing offspring. One may schematically think of a node as one neurone that fires and the offspring as the nodes receiving charges from the first node. Fig. \ref{Branch_fig} indicates the process. Every node produces $k$ offspring with the same independent probabilities $p_k, k=1,2,...$. The size of the generated tree, $S$, i.e. the total number of nodes produced by the root node, follows a distribution of the form\cite{Harris1963}
\begin{equation}
	P(S)\sim s^{-3/2}e^{-S/S_0(\mu)}\;\mbox{for $S\gg1$},
	\label{tree_distri}
\end{equation}
where the scale $S_0$ is determined by the average branching ration
\begin{equation}
\mu = \sum_{k=0}^\infty kp_k.	
\end{equation} 
When the branching ration $\mu$ becomes equal to one, the process is said to become critical. As $\mu=1$ is approach the scale $S_0$ diverges according to $S_0\sim (1-\mu)^{-2}$. This means, see Eq. (\ref{tree_distri}),  that the avalanche size distribution for the critical branching process is described by a power law with exponent of 3/2. 

The duration of the avalanche is given by the total number of generations $T$ until no new descendants of the root node are produced, see Fig. \ref{Branch_fig}. When the branching ration is brought to the critical value $\mu=1$ the distribution of durations also becomes a power law with $P(T)\sim T^{-2}$

How can uncorrelated events produce scale free power laws. Because the the variable $n(t)$ denoting the number of nodes in generation number $t$ after the initiation by the root will be correlated. Though the offspring production of any node is independent of any other node, the size of generation $t+1$ is clearly not independent of the size $n(t)$ of the previous generation. If there are many nodes in generation $t$, i.e. $n(t)$ large, the number of nodes in generation $t+1$ is more likely also to be large, than if $n(t)$ is small. But this dependence is not very profound.

The situation for a random walk signal $f(t)$ is similar. If $f(t+1)=f(t)+\Delta_t$ where $\Delta_t$ are random numbers drawn independently from the same distribution, the changes in $f(t+1)-f(t)$ are independent random numbers but the signal $f(t)$ is nevertheless characterised by by power laws. For example, consider the distribution of first return times $T$. The first return time is the time $T$ before $f(t+T)$ for the first time returns to the value $f(t)$ assumed at time $t$. This distribution  follows a power law $T^{-3/2}$. 

After having recapitulated that uncorrelated stochastic processes can produce power law avalanche distributions, the question is how relevant this is for the discussion about whether the brain exhibit critical behaviour and more specifically how relevant, and in what way, SOC is to brain dynamics. Let us first recall that the exponent 3/2 and 2 indeed have been reported for brain neuronal activity avalanches observed at very different scales. It is well known that at the level of individual neurones firing, Beggs and Plenz  \cite{BeggsPlenz2003} found the exponent 3/2 and 2 and the 3/2 exponent was also observed by Chialvo and collaborators for the distribution of avalanches sizes detected by fMRI\cite{Tagliazucchi2012}.

We should however keep in mind that accurate determination of the exponents is difficult and alternatives to power law behaviour has been suggested, see e.g. \cite{Wilting2019}. Our focus here will be slightly different, namely how approximate power law behaviour may still be interesting in terms of long range correlations.   

Even within the framework of SOC, the accurate values of these exponents and how they depend on the details of the brain activation has been the focus of detailed discussion, see e.g. \cite{Arcangelis2017} and references therein. Here we just want to stress that although the individual neuronal avalanches may exhibit exponents characteristic of an uncorrelated branching process with branching ratio equal to one, the waiting time distribution between the avalanches show that the initiation of the avalanches are correlated. It is well know that if neuronal avalanches were released with constant probability per time unit, i.e. as a Poisson process, the time between avalanches would be exponentially distributed. In contrast analysis shows that the waiting time is power law distributed\cite{Lombardi2012,Arcangelis2017}. These indicators of correlated brain activity are supplemented by the correlations in both space and time extracted from fMRI reported in \cite{Expert2011a} where power law decay as function of spatial separation were observed together with 1/f temporal power spectra indicating logarithmic decay of correlations in time. So indeed it is still relevant to understand better the aspects of the establishment through self-organisation of a correlated critical state.

Before we in the next section turn to this discussion, we will finish our brief overview by pointing to another indication of the relevance of the avalanche dynamics of SOC. In a series of work de Arcangelis and collaborators have developed SOC inspired models\cite{DeArcangelis2006}, which have been developed to allow analysis of the importance of inhibition\cite{Lombardi2012,Arcangelis2017}, modular structure\cite{Russo2014a}, the role of synaptic plasticity\cite{MichielsVanKessenich2016} and even address how avalanche dynamics may be able to support pattern recognition in a neuronal network\cite{MichielsVanKessenich2019}. This bulk of work certainly suggest that even beyond the question about criticality, SOC inspired modelling and analysis is highly relevant. 

\section{Burst dynamics and correlations}
Let us now turn to the question concerning in what sense self-organisation to criticality occurs and what different systems and models may share in this respect, which will lead us to suggesting that even if \textit{no perfect} tuning to a critical state in the sense of equilibrium statistical mechanics happens when judged from the scaling of the event distributions, interesting long range correlations may still play an important role and may also be operationally accessible. 

We will elaborate on the observation put forward by Chialvo and collaborators that the spatiotemporal statistics of the burst of activities in the brain as monitored by fMRI\cite{Chialvo2010,Tagliazucchi2012} exhibits remarkable commonalties with the observed activity of precipitation in the atmosphere\cite{Peters2006}. In turn the statistics of the brain activity and precipitation both behave very similar\cite{Palmieri2018} to the activity of the Forest Fire Model (FFM) of SOC\cite{Drossel_Schwabl1992}.

We need to describe the FFM briefly. The dynamics of the model can be thought of as a schematic representation of spreading across a population of some activity. The model considers a hyper-cubic lattice of dimension $d$, here we only discuss results for $d=2$ and use a square lattice of linear size $L$ containing $N=L^2$ sites. Each lattice site can be in three states: empty $E$, contain a tree $T$ or contain a tree on fire $F$. The basic dynamics consists of the following steps. Choose an empty site and plant a tree with probability $p$: $E\rightarrow T$. Choose one of the sites occupied by a tree at random and ignited it, i.e. turn it into a fire site, with probability $f$. A tree next to a site on fire becomes a fire site in the next time step: $T\rightarrow F$. A fire site becomes and empty site in the next time step: $F\rightarrow E$. The model has been studied extensively, for a recent simulation study see e.g. \cite{Palmieri2020} and references therein.   

The FFM is studied in the limit of $f\ll p$, i.e. very rare external ignition of fires. In this limit the salient aspects of the dynamics consists in the formation of connected clusters of tree sites. Since external ignition happens very rarely, a connected cluster is considered to burn down instantaneously on the time scale of the input of external energy represented by the external ignition. Simulations have shown that the distribution of sizes of burned-down clusters follows approximate power laws. But the detailed nature of of criticality and scale invariance in the model has been found to be unusual\cite{Grassberger2002,Pruessner2002}. It has for instance been suggested by Grassberger\cite{Grassberger2002} that the FFM model for large enough values of the ration $\theta=p/f$, and hence systems sizes big enough to accomodate the huge tree clusters generated, may exhibit the scaling known from percolation. This may be the  case but the analysis of \cite{Palmieri2020a} indicates that this scaling behaviour will not be observed until $\theta>10^{40}$, which to avoid that the simulations are influenced  by non-wanted  effects due to the finite size, will need totally unreachable system sizes, namely linear size of order $L\sim 10^{40}$\cite{Palmieri2020}. 

We conclude from this that the true asymptotic (system size going to infinity) behaviour is beyond reach at present, but hurry to say that the models behaviour for system sizes that can be studied by simulations appear indeed to be instructive as a way to gain insights on how SOC is relevant to real systems. 
 
\begin{figure}[!htbp]
        \centering\noindent
        \hspace{1.7cm}
       \includegraphics[width=13cm]{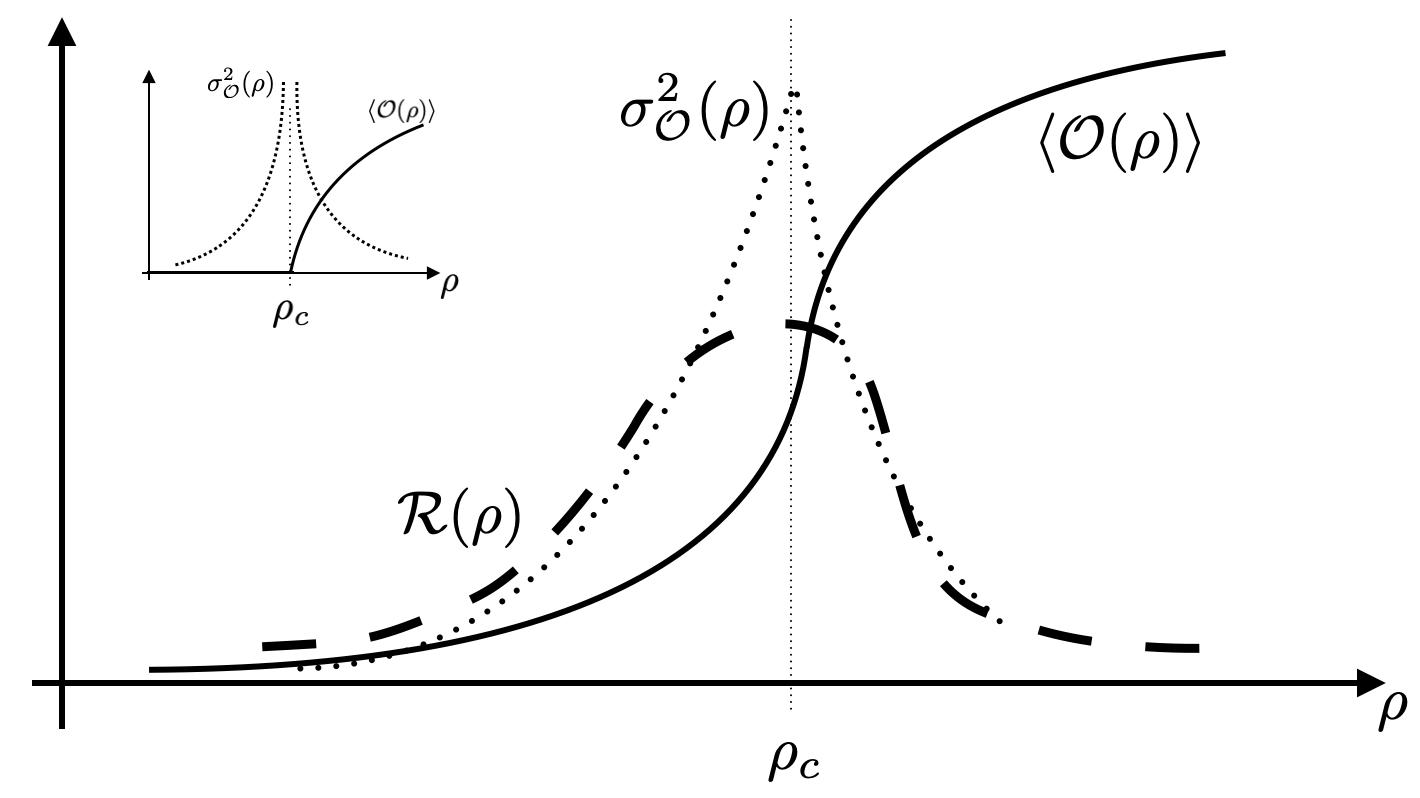}
       \caption{Generic sketch of a broad region of onset of behaviour reminiscent of the criticality familiar from equilibrium phase transitions. The three curves indicate the following: The solid the order parameter, the long dashed line the resident time and the dotted line the variance (in equilibrium systems equivalent to the susceptibility). In a critical equilibrium phase transition (depicted in the insert), such as the Ising model and magnetic systems, the order parameter is strictly zero on one side of the transition for then to become non-zero at a sharp value, the critical value, see $\rho_c$ and the susceptibility diverges at $\rho_c$. The residence time does not have an equivalent in equilibrium systems because the control parameter $\rho$ is fixed externally. But only one value of $\rho$ is relates to each equilibrium state.}
       \label{Generic_onset}
\end{figure}

Let us briefly sketch the situation. Peters and Neelin\cite{Peters2006} observed that when the amount of precipitation is plotted agains the humidity of the atmosphere, a sharp onset is seen in a narrow range of humidities. Fig. \ref{Generic_onset} shows a depiction of the behaviour relevant to precipitation, brain activity and activity in the FFM model. 

The figure summarises the behaviour presented in four figures in the literature. Firstly, Fig. 1  in \cite{Peters2006} that plot the amount of precipitation and its variance as function of the amount of vapour in the atmosphere. The time the atmosphere is found to spend at a certain vapour level is presented in Fig. 3 in \cite{Peters2006}. This time is called the residence time. Secondly, Fig. 3E in \cite{Tagliazucchi2012} that shows the normalised size of the largest size of the cluster of active fMRI voxels (above a certain threshold) as function of the total number of active voxels, its variance and the residence time. Finally, a similar analysis was presented in Fig. 1 of \cite{Palmieri2018} that contains the largest tree cluster, its variance and the residence time plotted against the number of trees (cnsidered to be the sites able to generate activity).

In all three cases it was observed that the so-called order parameter, i.e the parameter that distinguish where the system is relative to the critical state, pickups from very low value over a relatively narrow region of values of the control parameter. The order parameter for the three considered cases are: amount of precipitation, size of connected cluster of active voxels and size of burnt down clusters. The control parameter for the three cases are: amount of vapour in the atmosphere, number of active voxels and number of sites occupied by trees. In all three cases approximative power laws for the event distribution (the size of precipitation events, the size of the largest cluster of active voxels, the size of the largest burnt down cluster of tress, respectively) are observed at some value of the control parameter at about the middle of the region where the order parameter picks up and where the variance peaks. This means that in all three cases we observe a spread out version of what would appear as a sharp transition in critical equilibrium phase transitions, see insert in Fig. \ref{Generic_onset}. 

We want to stress two important points. First that these three entirely different systems share remarkable similarities of how they organise themselves in a way so they are most frequently found in a region exhibiting aspects of an underlying nearby critical transition. The other point we want to emphasise is that despite both the brain and the atmosphere being very large systems, they share behaviour with the \textit{finite size} properties of the FFM. This suggests that in contrast to equilibrium critical phenomena, for which we know it is crucial to understand the asymptotic behaviour in the limit of diverging system size, in the case of manifestations of the avalanche dynamics associated with SOC, the finite system size behaviour is of particular interest. This implies on the other hand that rather than self-organisation to one sharply defined critical point, the self-organisation of relevance to applications consists of a tendency for the dynamics to bring itself to a certain region of near critical behaviour.

The question then remains what makes this region near some kind of criticality interesting. We therefore return to correlations and recall that the power  laws and diverging sensitivity (susceptibility) of the equilibrium critical state all are a consequence of long range algebraic correlations. Is there any evidence that the region, where our three systems exhibit a peak in the residence time, is associated with correlations behaving as described by Eq. (\ref{Corr_func_2}) with a correlation length $\xi$ large enough to make the slow power law decay $r^{-a}$ able to essentially determine the behaviour. 

We do not know yet the answer for the atmosphere. The fMRI study of the BOLD signal correlations in \cite{Expert2011a} suggests this is the case for the brain and recent studies of the FFM confirm that despite of the puzzling scaling observed for the event distributions, spatial correlations are long range\cite{Palmieri2020a}. In this approach the correlation length is considered to be a stochastic variable. For each instantaneous configuration of the considered system, the correlation function  in Eq. (\ref{Corr_func_1}) is fitted to the functional form in Eq. (\ref{Corr_func_2}) in order to determine the correlation length $\xi$ describing this specific configuration. This approach differs from the usual way the correlation length is determined in studies of equilibrium system. Simulations usually first average the correlation function in Eq. (\ref{Corr_func_1}) over all the generated configurations and then afterwards extract the correlation length from a fit to this averaged function. The new method, which focus on the correlation length as a characteristic of each individual configuration, was shown in \cite{Palmieri2020a} to reproduce the known equilibrium behaviour for the two prototype  models: the Ising model and the 2d-XY model. And more interesting, the method produces new understanding of correlations in the FFM.

For the FFM model, the variable $S_i$ is taken to be $S_i=1$ for tree sites and $S_i=0$ for empty sites. Fire sites are ignored since fires remove connected clusters in a single step on the time scale given by $p$ the tree growth probability. As growth and fire dynamics is simulated, the distribution, $P(\xi)$, of the correlation lengths for instantaneous configurations  is sampled.  Simulations found that for system sizes up to $L=3000$, the distribution $P(\xi)$ is well described by finite system size scaling meaning that $P(\xi)$ for different values of $L$ can be collapsed to one ``universal'' function when the activity per site given by $\theta/L^2$ is kept fixed as $L$ is varied. The dependence on system size indicates that the average correlation length extracted from the distribution $P(\xi)$ depends on system size as $\langle\xi\rangle\propto L^{1.1}$. This suggests that as $L$ increases correlations will become of significant long range when the drive given by the fire probability $f$ is well separated from the growth probability $p$, i.e. when the external drive is slow.

\section{Summary and Discussion}
We have notice that the exact tuning to a critical point first anticipated by SOC seems not to happen in rain, brain and the Forest Fire Model. This does not leave the framework of SOC irrelevant. On the contrary self-organisation does occur though to a region near the rise of an order parameter. Moreover, studies of the FFM model suggest that although power laws of event distributions may only be approximate the region where these three systems spend most time is nevertheless associated with very extended spatial correlations. 

The similar behaviour shared between so different systems indicates that the basic ingredients of dynamics driven by  load, spreading and relaxation will tend to organise towards configurations poised near some kind of onset of (correlated) percolation. The buildup of spatially extended structures, which then abruptly collapse through release of precipitation, neuronal firing or fires respectively in the three cases discussed, keeps turning ``over critical'' configuration into under ``under critical'' ones. 

Simulations of the FFM show that for parameter regimes manageable the region of preferentially visited configurations, i.e. the region around the peak in the residence time distribution, broadens as the separation between the two time scales given by loading (tree growth) and triggering of release of energy (externally induced fires), i.e. $\theta=p/f$ increases. The width of the peak of the residence time in the FFM behaves approximately like $\sigma\sim\theta^{0.1}$, see \cite{Palmieri2020}. So driving the model towards more extended power laws and longer correlation lengths also broadens the range of visits to both under and over critical configurations. 

The conclusion is that the load and release dynamics does not manage to fine tune to a critical point but to a region of extended correlations.

This has been suggested before, see for example \cite{Bonachela2009,Bonachela2010,Wilting2019}, but here we emphasise that though strict criticality is not achieved correlations are sufficiently extensive to make load and relaxation dynamics in the three completely different systems rain, brain and FFM share remarkable similarities. 

It is often suggested that the reason the brain operates at or near a critical point is the hypersensitivity of this state, indicated by the divergence of the susceptibility.  This seems certainly very reasonable, but there may even be reasons why the brain does not sit exactly in a critical state. As suggested by Lizier and collaborators\cite{Li2019} operating in the region across the critical point may have a computational advantages for the brain in terms of combining high data storage capability in the sub-critical region with increased information transfer in the super-critical region. So after all, lack of exact tuning to criticality may in the end be seen as a bonus.

\section{Acknowledgments}
It is a great pleasure to acknowledge stimulating interaction spanning many years with Dante Chialvo and Lucilla de Arcangelis and the recent very fruitful collaboration with Lorenzo Palmieri. I am grateful to Hardik Rajpal for pointing me to the work by Lizier and collaborators.

\section{References} 
\bibliographystyle{unsrt}

\end{document}